\documentclass[runningheads]{llncs}

\usepackage{xcolor}

\usepackage{graphicx}
\usepackage{comment}
\usepackage{amssymb}
\usepackage[many]{tcolorbox}
\newtcolorbox{mybox}{colback=blue!10}

\usepackage{parskip}
\usepackage{subfigure}
\usepackage{stfloats}
\usepackage{url}
\usepackage{graphicx}
\usepackage{caption}
\usepackage{svg}
\usepackage{changepage}
\usepackage{booktabs}

\renewcommand\labelitemi{\normalfont\bfseries --}
\renewcommand\labelitemii{$\tiny\bullet$}

\begin{document}

\title{Practitioner Paper: Decoding Intellectual Property: Acoustic and Magnetic Side-channel Attack on a 3D Printer}
\titlerunning{}
%
\author{Amirhossein Jamarani \orcidID{0009-0009-0692-9076}\and  Yazhou Tu$^\dag$ \orcidID{0000-0001-7640-1829} \and
Xiali Hei \orcidID{0000-0002-2438-5430} }

\authorrunning{A. Jamarani et al.}

\institute{University of Louisiana at Lafayette, Lafayette, LA, 70504, USA \\
$^\dag$Auburn University, Auburn, Alabama 36849, USA}


\titlerunning{Practitioner Paper: Decoding Intellectual Property}
\maketitle              

\begin{abstract}
 The widespread accessibility and ease of use of additive manufacturing (AM), widely recognized as 3D printing, has put Intellectual Property (IP) at great risk of theft. As 3D printers emit acoustic and magnetic signals while printing, the signals can be captured and analyzed using a smartphone for the purpose of IP attack. This is an instance of physical-to-cyber exploitation, as there is no direct contact with the 3D printer. Although cyber vulnerabilities in 3D printers are becoming more apparent, the methods for protecting IPs are yet to be fully investigated. The threat scenarios in previous works have mainly rested on advanced recording devices for data collection and entailed placing the device very close to the 3D printer. However, our work demonstrates the feasibility of reconstructing G-codes by performing side-channel attacks on a 3D printer using a smartphone from greater distances. By training models using Gradient Boosted Decision Trees, our prediction results for each axial movement, stepper, nozzle, and rotor speed achieve high accuracy, with a mean of 98.80\%, without any intrusiveness. We effectively deploy the model in a real-world examination, achieving a Mean Tendency Error (MTE) of 4.47\% on a plain G-code design.

\keywords{3D Printer  \and Side-channel Attack \and G-code Reconstruction \and  Physical-to-cyber Attack \and Intellectual Property.}
\end{abstract}

\section{Introduction}


The emergence of 3D printers dates back to the 1980s. The foundational concept laid the groundwork for the development of modern 3D printers. Today, as additive manufacturing (AM) systems grow and become prevalent globally for various purposes ranging from industrial usage to healthcare \cite{awad20183d,dodziuk2016applications}, biomedical \cite{lai20213d}, aviation \cite{kalender2019additive}, energy, and consumer products \cite{brooks20143d,kalender2019additive,lai20213d} the importance of keeping the Intellectual Property (IP) of these systems safe and secure has significantly increased. The ease of working with 3D printers and their efficiency have led industries to produce both high-tech and regular goods using G-code because it is cost-effective, flexible, accessible, and reliable. As predicted in \cite{10.1145/2976749.2978300}, the global revenue of 3D printers reached over 20.2 billion dollars in 2021. It is calculated to produce revenue of 162.7 billion dollars by 2030 at a Compound Annual Growth Rate (CAGR) of 23.6\% \cite{ree2024critical,3DMarket}. The concept of cyber-physical attacks on 3D printers has been explored since 2014, with research monitoring and analyzing these attacks \cite{sturm2017cyber,moore2016vulnerability}. 

3D printers emit acoustic signals and generate magnetic fields, raising the question: \textit{could these emissions be recorded by smartphones and used to reconstruct the G-code?} Since 3D printers have digital acoustic signatures \cite{belikovetsky2018digital} for each movement and also generate magnetic fields, 
attackers could utilize a smartphone's built-in sensors including the microphone to capture these data without physically contacting the printer.
By analyzing these recordings, attackers can potentially reconstruct the original G-code and commit successful IP theft \cite{holland2018intellectual}. Moreover, advancements in smartphone sensors have made it increasingly easier for attackers to accurately and discreetly collect data, allowing them to access IPs without being physically close to the 3D printer.

In this paper, we conducted a real test-bed attack on a 3D printer by analyzing multiple side channels emitted during the printing process using a smartphone for data collection. We examined the relationship between G-code commands and IP through distinct movements of the 3D printer: vertical or horizontal movements (left and right or up and down), header, and strata movements. Using the side-channel data, we trained Gradient Boosted Decision Trees on each movement. Subsequently, we applied the Side-Channel Reconstruction of the G-code (SCReG) technique, which utilizes acoustic and magnetic emissions generated by a 3D printer to infer and reconstruct the original G-code instructions. By collecting data through a smartphone’s sensors, this method employs machine learning models to analyze the side-channel information and predict the printer's movements. The reconstructed G-code can then be used to replicate the printer’s operations, potentially bypass the security measures, and access the IP coded in the printing process. This approach demonstrates the feasibility of reverse-engineering 3D printer instructions by only monitoring side channels. 

The Mean Tendency Error (MTE) of our research attained the lowest percentage of 4.47\%, highlighting that the reconstructed G-code and the reverse-engineered printed object were very similar to the initial object printed by the user. The accuracy of our models varied for each movement of the 3D printer discussed in Section 5. Our study introduces an approach to reconstructing G-code commands using machine learning algorithms with minimal MTE and inaccuracy. This research illuminates previously unexplored areas of side-channel analysis, such as setting up the smartphone for data collection at further distances of the 3D printer and the non-intrusiveness nature of the attack. Also, it utilizes feature extraction from acoustic and magnetic data to achieve more accuracy. The key contributions of our study are outlined below:
\let\labelitemi\labelitemii

\begin{itemize}
  \item We fully analyze the side channels produced by the 3D printer in different axes of nozzle movements
  and train a model to predict the movements.
  \item We reconstruct the G-code commands with the usage of a machine learning algorithm, Gradient Boosted Decision Trees,  with high accuracy and low MTE. We illuminate the procedure of how to use a smartphone to collect data from 3D printer in an effective way.
  \item This study provides a technical and comprehensive taxonomy of the attack model. We discuss/identify open challenges and future trends in side-channel attacks on 3D printers.
  
\end{itemize}

The remaining parts of this study are organized as follows: we survey the related literature in Section 2. We provide a background and operational mechanism of AMs and 3D printers with the commonly used third-party tools to interpret the design to the 3D printer in Section 3. Then, we introduce a threat model and examine the side channels on the 3D printer in Section 4. We present the trained acoustic and magnetic models to predict movements in Section 5. In the following section, Section 6, we showcase our model and results in a real-world test-bed. Finally, Section 7 concludes the study.

\section{Related Work}
Extensive research has been conducted in the domain of either acoustic or magnetic side channels on 3D printers. However, less focus has been on the area where both side channels are employed. This section reviews the related studies in the field. The comparison of the works that are most related to ours is discussed in Section 6.1.

In \cite{song2016my}, the authors' approach to reconstruct the G-code reached an MTE score of 5.87\% by implementing a five-layer operational analysis consisting of Layer movement, which modeled to diagnose if the 3D printer has Z-axis movement (changing to another layer) or if it is in X-Y plane. Then, header movement was examined to detect if the nozzle was printing or if it was just changing the position to align with no material extrusion. The subsequent layer was axial movement to discern if the nozzle was moving in X-axis or Y-axis. The last two layers were designed to spot if the nozzle is in X-axis movement or if it is moving in X-left or X-right. The same was designed for the Y-axis movement to perceive if the nozzle is moving in Y-up or Y-down. Our study aligns with the work done in \cite{song2016my}; however, the main differences are the distance the smartphone was placed to the 3D printer, the algorithm used, and applying different feature extractions to have more robust and clean data.



Authors in \cite{chhetri2016novel} completed a thesis on cyber-physical attacks in additive manufacturing systems. 
They updated that physical-to-cyber attacks exploit manifestations of cyber-domain information through physical actions like motion and temperature changes, leaking confidential data via side-channels such as acoustic, thermal, and power. Their thesis investigated how acoustic side-channels can be used to gain the confidentiality of AMs, such as 3D printers, by reconstructing the G-code commands and IPs. Their attack model, including digital signal processing and machine learning algorithms, restored test objects with 78.35\% axis prediction accuracy and 17.82\% length prediction error.

In cyber-physical domains \cite{chhetri2017confidentiality}, the integration of side channels makes the systems vulnerable to attacks, exploiting information, like thermal, acoustic, and power conduits, to extract data without any disturbance to the functional system. As a case study, the authors implemented an FDM-based model on 3D printers, depicting the fact that how acoustic data can represent information about what is being printed on the 3D printer. With a model of an attack and the usage of machine learning methods, they reconstructed G-code to access IPs stored in the cyber domain. Their method gained an average axis prediction accuracy of 86\% and an average length prediction error of 11.11\% on different simple objects.

Authors in \cite{Faruque2016AcousticSA} only deployed acoustic side-channel attack on a 3D printer. The concept was faster motor rotation (higher speed), which results in higher amplitude and frequency sounds, so by collecting only acoustic data and training a model (regression model), they could access the G-code. Their attack methodology was comprised of two phases: training and attack itself. During the training phase, they recorded audio signals, pre-processed, and examined feature extractions in time and frequency domains. Then, features were mapped to corresponding G-codes. Finally, regression and classification models were trained. For the attack phase, they collected audio frame data and pre-processed it as similar to what was done in the training phase. Features were passed to the trained models, and the Predicted data was used to reconstruct the G-code.

For their classification model, the authors \cite{Faruque2016AcousticSA} introduced four sections, naming phi 1 to phi 4, each classifying in the Z-axis (Z) and no movement in the Z-axis (-Z), one-dimensional (1D), and two-dimensional (2D) movement. If the movement is 1D, this classifier determines whether it is along the X-axis or Y-axis, and: If the movement is 2D and along the XY axes, this classifier determines whether the X and Y motors are moving at the same speed or different speeds. The attack model reconstructed a square (simple shape) with classification accuracy reaching 98.55\% and a Mean Absolute Percentage Error (MAPE) of 3.13\% after post-processing. The authors in \cite{YAMPOLSKIY201658} looked at 3D printers as a weapon. They explored potential risks aligned with the malicious intent of using 3D printers. They also pointed out that by only having minor modifications on the IP and the physical properties of the 3D printers, the overall object at the end can be turned into dangerous items. The paper offered different taxonomies covering possible types of attack scenarios and discussed the weaponizing scenarios based on each attack model. 

In another aspect of securing G-codes and preventing any malicious modifications, the two studies in \cite{10.1115/1.4063859,10.1115/1.4048966} proposed an approach for encrypting the G-code at its initial level once it is out of Stereo-Lithography (STL) file for being sliced layer-by-layer. In \cite{10.1115/1.4063859}, the main purpose was to implement an approach to protect sensor data in cyber-enabled advanced manufacturing systems from cyber-physical attacks, in specific terms of unauthorized access and malicious tampering. However, in \cite{10.1115/1.4048966}, the authors mainly focused on blockchain-based G-Code storage and asymmetry encryption of the row by row of the G-code to provide a secure path for sender and receiver. They came up with two case scenarios of the attack: First, unintended design modifications, and second on intellectual property theft where unauthorized access to the G-code allows the attacker to reproduce the product without the owner's permission. 

The attack vectors on infrared (IR) thermography, used for quality control in metal additive manufacturing, was studied in \cite{10.1145/3098954.3107011}. The research diagnosed each possible attack scenario, such as manipulating calibration data and compromising thermal cameras, which can lead to defects like increased porosity or lack of fusion in the final product. The research highlighted the differences between open-loop and closed-loop systems to demonstrate the vulnerability of the 3D printer that while both can be achieved, the effects on part quality may change depending on the system's configuration. In \cite{BRANDMAN2020202}, the aim was to secure AMs from cyber-physical attacks by identifying a physical hash method. This method uses a quick read code that encodes a hash of process parameters and toolpaths, ensuring that any deviations during manufacturing are detected in a synchronized manner of the time. The research contributes to enhancing the security and quality assurance of AM processes by integrating this physical hash with side-channel monitoring systems.

IP protection in additive layer manufacturing (ALM) was explored in \cite{10.1145/2689702.2689709}. The authors introduced an outsourcing model for ALM that aimed to address the limitations of traditional outsourcing methods for the purpose of securing IPs. Authors in \cite{STURM2017154} also did similar work to the research done in ALM. They explored potential cyber-physical attack vectors within the AMs process chain. The research showcased that the current detection methods, such as machine operators, virus-checking tools, and STL validation software, are insufficient for identifying sophisticated cyber attacks.

\section{Background and Operational Mechanism}
\begin{figure}[h]
    \centering
    \includegraphics[width=0.90\linewidth]{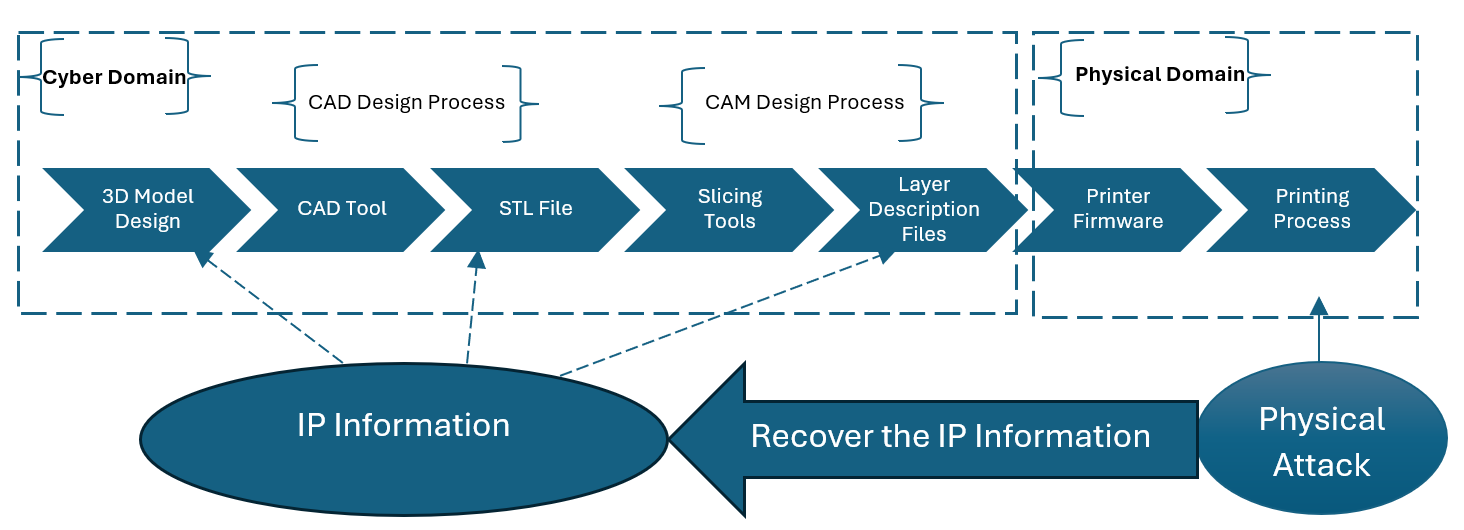}
    \vspace{-3mm}\caption{Lifecycle of AMs under physical attack} \vspace{-3mm}
    \label{fig1}
\end{figure}

As depicted in Figure \ref{fig1}, the typical life cycle of AM involves several stages. Users begin by designing a 3D object using design tools such as Fusion, Blender, or Sketchup \cite{chopra2012introduction}. After designing, a STL file is generated through Computer-Aided Design (CAD). Subsequently, Computer-Aided Manufacturing (CAM) tools are used to produce slicing instructions and layer description files, such as G-code. 

IP protection in AM is critical due to the risk of theft and unauthorized reproduction. Different implementations have been proposed to secure IPs, including digital watermarking, encryption \cite{10.1115/1.4063859}, and blockchain-based solutions \cite{10.1115/1.4048966}. In recent years, block chaining the G-codes has emerged as an effective resolution for securing the entire lifecycle of AM products by providing a tamper-proof record of all modifications and operations performed on the STL and G-code.

For production purposes, the components of the 3D printer, such as motors, steppers, nozzle, fan, and extruder, are commanded to run the lines specified to them by the G-code or the STL file. In this step, the printing process of an object happens. As known, the 3D printers emit acoustic sound and generate magnetic field while printing \cite{10.1145/2976749.2978300}; here, the attacker can take advantage of these data by recording them. Having analyzed and trained the data by machine learning or other applicable methods, the attacker recovers the IP data from the physical attack, which will lead him to gain full access to the IP information that was initially generated by the designer of the 3D object.  

\begin{figure}[h]
    \centering
    \includegraphics[width=0.8\linewidth]{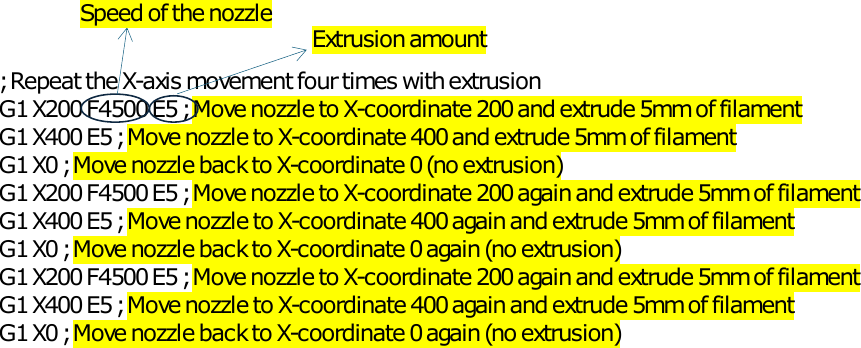}
    \vspace{-3mm}\caption{Specific G-code commands for only moving the nozzle on the x-axis} \vspace{-3mm}
    \label{fig:your-image}
\end{figure}

The major Segments of a 3D printer are actuators and motors. These motors control the nuance movements of the printer's parts, including the extrusion nozzle and the print bed, along the X, Y, and Z axes. The main responsibility of an extruder is to feed the filament into a nozzle, where it is melted by the pre-heated nozzle and released onto the pre-heated bed. The extruder contains a stepper motor that drives the filament through the nozzle. As the filament is extruded, the stepper motors move the nozzle according to the G-code instructions, constructing the object layer by layer. The pulse Width Modulation (PWM) technique is applied in each 3D printer to balance the power and current rushed to the motors and stators. 
The bed is the surface on which the object is printed. The stepper motors also move the build platform up and down along the Z-axis as new layers are added. Some 3D printers include auto-leveling sensors to ensure the print bed is level before printing begins, improving the quality and accuracy of the final object \cite{saggiomo20223d}.

Figure 2 represents a G-code for only moving the nozzle in the X-axis from right to left and reverse. Each row specifies the nozzle speed and movement coordinates. Also, the amount of filament per spot is controlled and can be modified based on each object's design. These commands are considered the IPs of a 3D printer, each of which can be reconciled with a distinct emission of sound and magnetic field.

\begin{figure}
    \centering
    \includegraphics[width=0.60\linewidth]{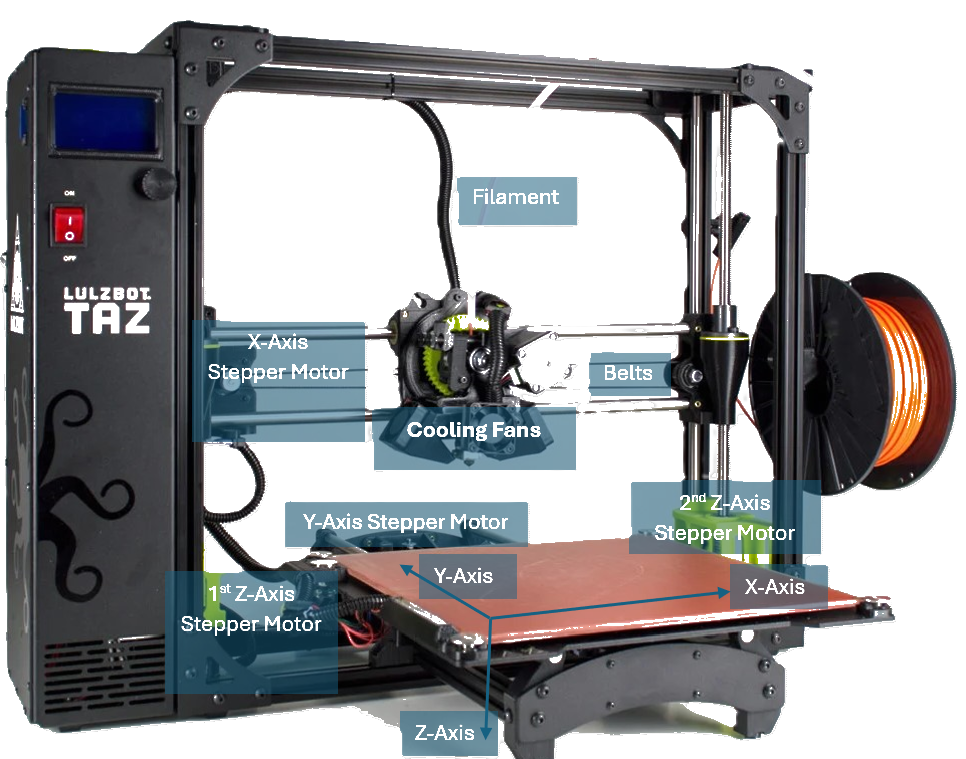}
    \vspace{-3mm}\caption{LULZBOT TAZ 3D printer with a heating nozzle and platform.} \vspace{-4mm}
    \label{fig:enter-label}
\end{figure}

A common 3D printer, LULZBOT TAZ \cite{WinNT}, is shown in Figure 3. It consists of a heating platform, which only moves on a Y-directional axis with the help of the Y-Axis stepper motor. In this particular 3D printer, there are two Z-axis stepper motors, each at the very edge of the 3D printer's platform. These only move top and bottom to adjust the nozzle for different printing process layers. The nozzle, the cooling fans, and the X-axis stepper motor are connected to the z-axis mover. This gives the 3D printer flexibility to align and print at each single spot on the platform. The X-axis stepper motor moves the nozzle only horizontally to either left or right, with the stepper connected to the transitional belts in parallel.  The Y-axis stepper motor sole positions the platform and moves vertically, either backward or forward.
The printing process generally involves heating the nozzle to change the material from a solid to a semi-solid stage. Temperature regulation is gained through cooling fans and heaters connected to the nozzle and the platform. The motors and their actuation systems manage the entire printing operation as those run the commands on the G-code or STL files.

\section{Threat Model and Side-channels}
As 3D printers produce acoustic sound and generate magnetic field while printing \cite{bilal2017review,10.1145/2976749.2978300}, attackers can reach the initial G-code by recording the acoustic and magnetic data either separately or concurrently using a smartphone application. For the simultaneous recording of magnetic field and acoustic sound, there are applications, such as \cite{toolbox}, where the users can record acoustic data and magnetic data at the same time. The output will be in a comma-separated value file, where can easily be analyzed and labeled for training a model. 

A potential attack model has been depicted in Figure 4. In this case scenario, the attacker places a smartphone near the 3D printer to record emitted data, both acoustic and magnetic. 
Then, he uses pre-trained learning algorithms to reconstruct the G-code. This allows the attacker to acquire the IP of the object being printed without even touching the 3D printer or disrupting the printing process.

\begin{figure}[h]
    \centering
    \includegraphics[width=0.9\linewidth]{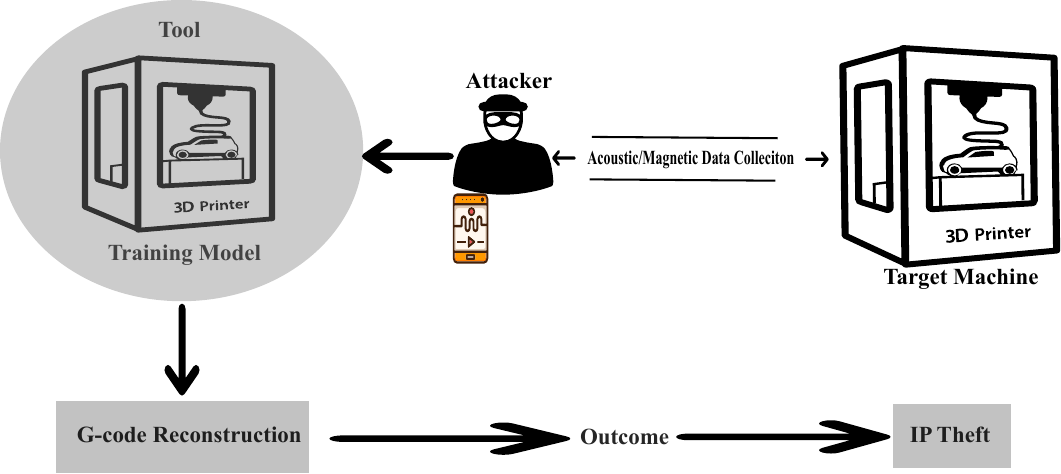}
    \caption{Threat model} \vspace{-3mm}
    \label{fig:your-image}
\end{figure}

\subsection{Acoustic Side-channel}

\begin{figure}[h]
    \centering
    \includegraphics[width=0.9\linewidth]{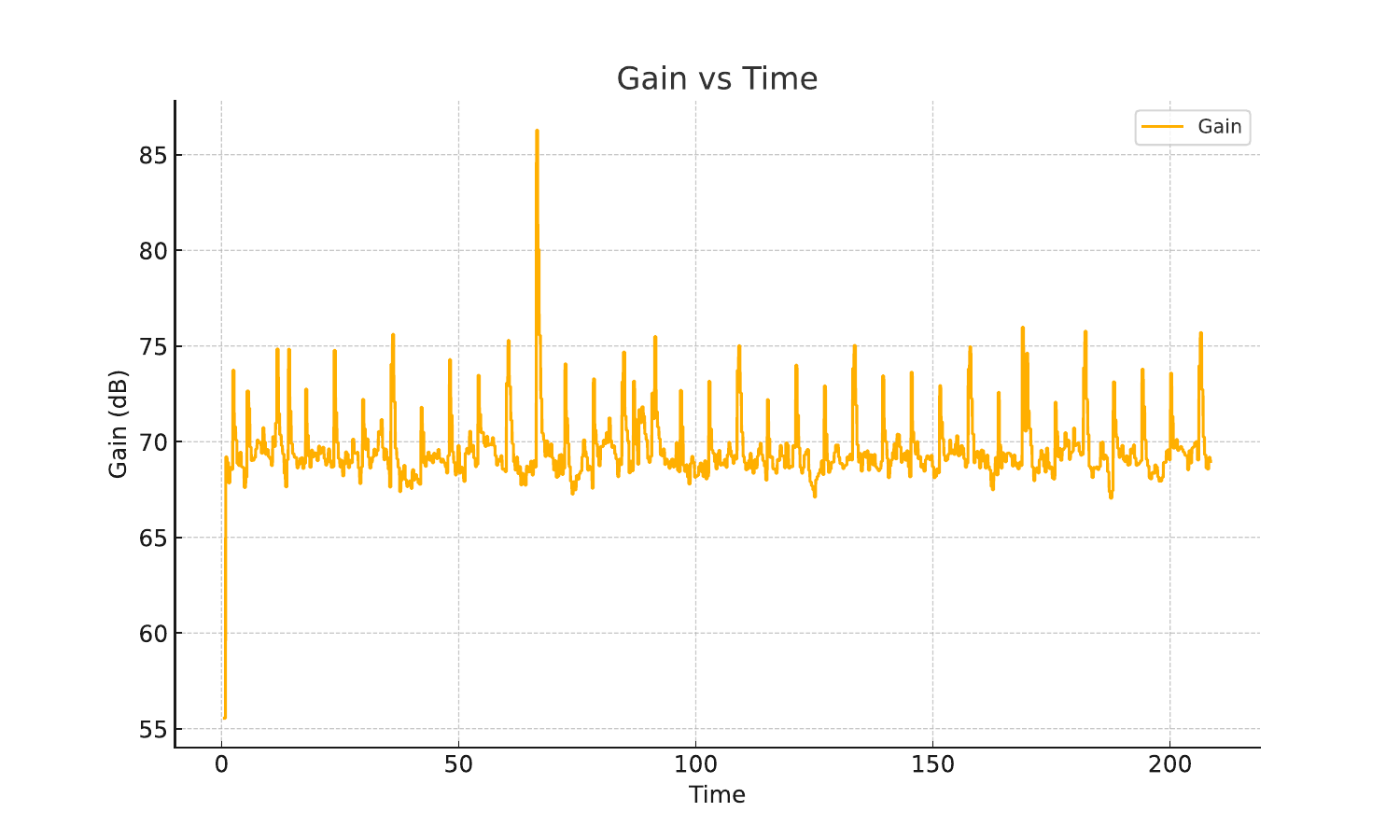}
    \caption{Gain vs. time for x-axis movements} \vspace{-3mm}
    \label{fig:your-image}
\end{figure}

By recording the acoustic data emitted from the 3D printer's nozzle, extrusion, layer, header, axial, and directional movements, the IP reconstruction will be feasible by an attacker. As depicted in Figure 5, the acoustic data of a 3D printer \cite{WinNT} was collected with a microphone of a smartphone, Samsung Galaxy S22 plus \cite{Samsung}. As it is apparent in Figure 5, there are trends at which level of Decibel (dB) the sound is emitted. Passing over time (in seconds), a continuous cycle of dBs ranges at certain points. 
In such case, to have better and more accurate data, feature extraction techniques \cite{lambrou1998classification,dhanalakshmi2009classification} were applied to the dataset. This helps to remove any background or unwanted noise, which leads to precise data.  

Applying Formula (1), Zero-Crossing Rate (ZCR), the rate at which a signal changes from positive to negative or back is monitored. Here, $x[n]$ is the signal at sample $n$ and $N$ is the total number of samples.

\begin{equation}
\text{ZCR} = \frac{1}{N-1} \sum_{n=1}^{N-1} \mathbb{1} \{ x[n] \cdot x[n-1] < 0 \}
\label{eq:ZCR}
\end{equation}

We applied the short-time energy Formula (2), which filters out low-energy segments that are not relevant to the overall acoustic data collection of the 3D printer. This improved the total performance of processing stages by spotting only the significant parts of the signal.

\begin{equation}
\text{STE}[n] = \sum_{m=0}^{N-1} x^2[n - m]
\end{equation}
where \( x[n] \) is the audio signal and \( N \) is the window length. 
We applied Formula (3), Root Mean Square (RMS), as it is less sensitive to short-term signal fluctuations. This makes RMS a more robust metric for evaluating the total level of our noisy signals.

\begin{equation}
\text{RMS} = \sqrt{\frac{1}{N} \sum_{n=0}^{N-1} x^2(n)}
\end{equation}

As we implemented a machine learning algorithm to train the data, it was necessary to use the spectral centroid, Formula (4), as it classifies different spectrums of sounds. With the help of spectral centroid, our model could detect the background noise and the noise emitted out of the 3D printer.

\begin{equation}
\text{Spectral Centroid} = \frac{\sum_{k=0}^{N-1} f(k) \cdot |X(k)|}{\sum_{k=0}^{N-1} |X(k)|}
\end{equation}
where \( f(k) \) is the frequency bin and \( X(k) \) is the magnitude of the Fourier Transform.

Spectral Bandwidth in Formula (5) was used to assist the model in understanding the spectral characteristics of a signal and applying necessary modifications.

\begin{equation}
\text{Spectral Bandwidth} = \sqrt{\frac{\sum_{k=0}^{N-1} (f(k) - C)^2 \cdot |X(k)|}{\sum_{k=0}^{N-1} |X(k)|}}
\end{equation}
where \( C \) is the spectral centroid.

Lastly, we used Gaussian filter to smooth our data in Formula (6).

\begin{equation}
G(x) = \frac{1}{\sqrt{2 \pi \sigma^2}} \exp \left( -\frac{x^2}{2 \sigma^2} \right)
\end{equation}
where \( x \) is the distance from the center of the filter, and \( \sigma \) is the standard deviation.

\begin{figure}[h]
    \centering
    \includegraphics[width=0.9\linewidth]{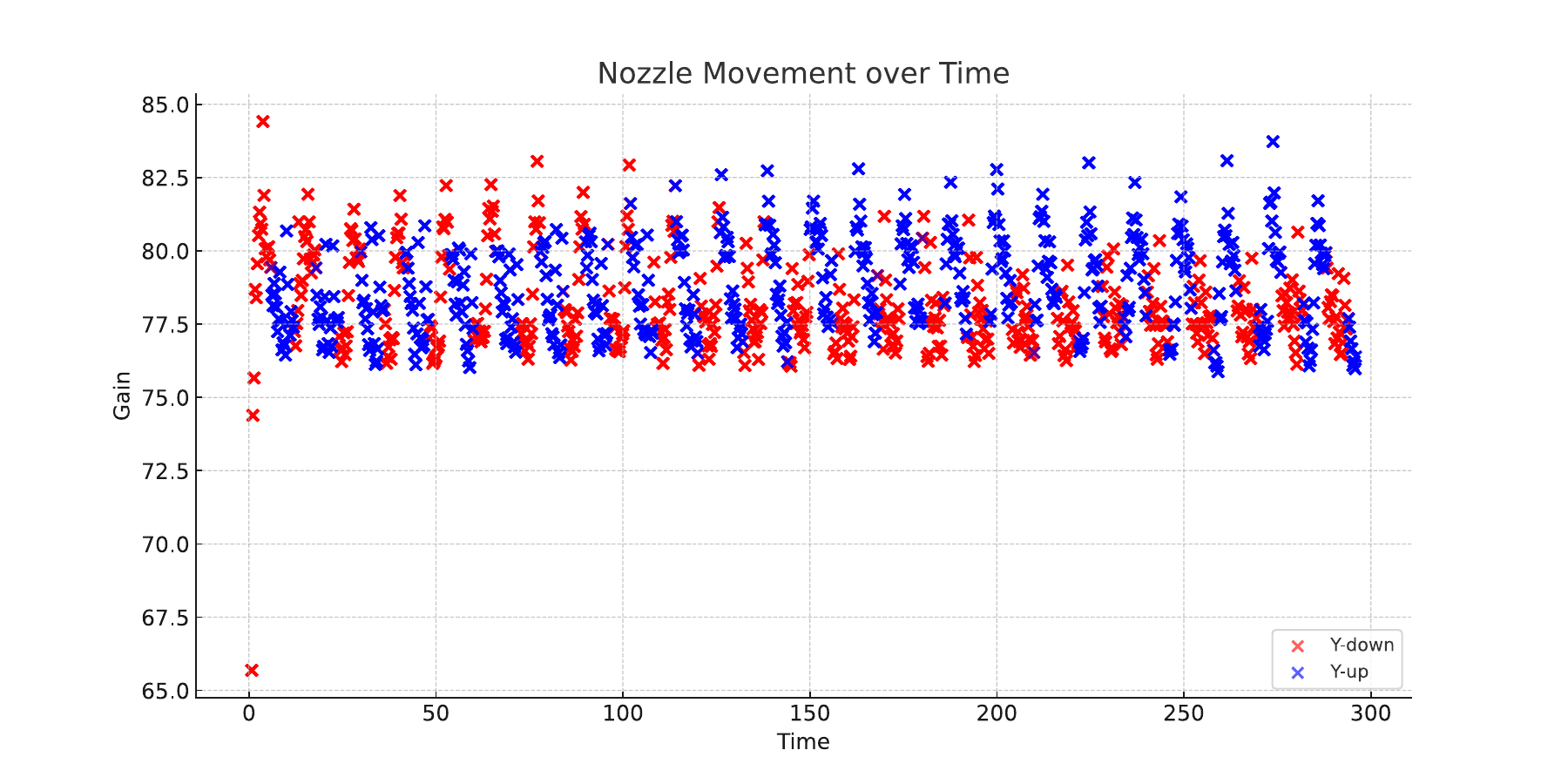}
    \caption{Trained model while the nozzle moving on y-axis backward and onward without extrusion} \vspace{-3mm}
    \label{fig:your-image}
\end{figure}

We mainly applied Mel-Frequency Cepstral Coefficients (MFCCs), Formula (7), to provide a relatively tight Portrayal of the spectral data of an acoustic signal. By analyzing the key information in a small number of coefficients, MFCCs greatly reduce the dimensionality of the data, making the machine-learning models to train the model easier and more classified.
MFCCs are derived as follows:

\begin{equation}
\text{MFCC}(n) = \sum_{m=1}^{M} \log(S(m)) \cos\left[n(m-0.5)\frac{\pi}{M}\right]
\end{equation}
Mathematically, if \( S(m) \) is the mel spectrum.

By applying all the feature extractions above to our acoustic data, we successfully trained a model to detect the movement of the nozzle in the y-axis, as illustrated in Figure 6. As it is apparent, the model predicted the movement with a high rate of accuracy with only some points missing due to the initial sound emitted by the 3D while it starts to initiate the G-code commands. Blue crosses are highlighted when the nozzle moves in Y-up, and red ones specify the Y-down movement of the nozzle.

\subsection{Magnetic Side-channel}
\begin{figure}[h]
    \centering
    \includegraphics[width=\linewidth]{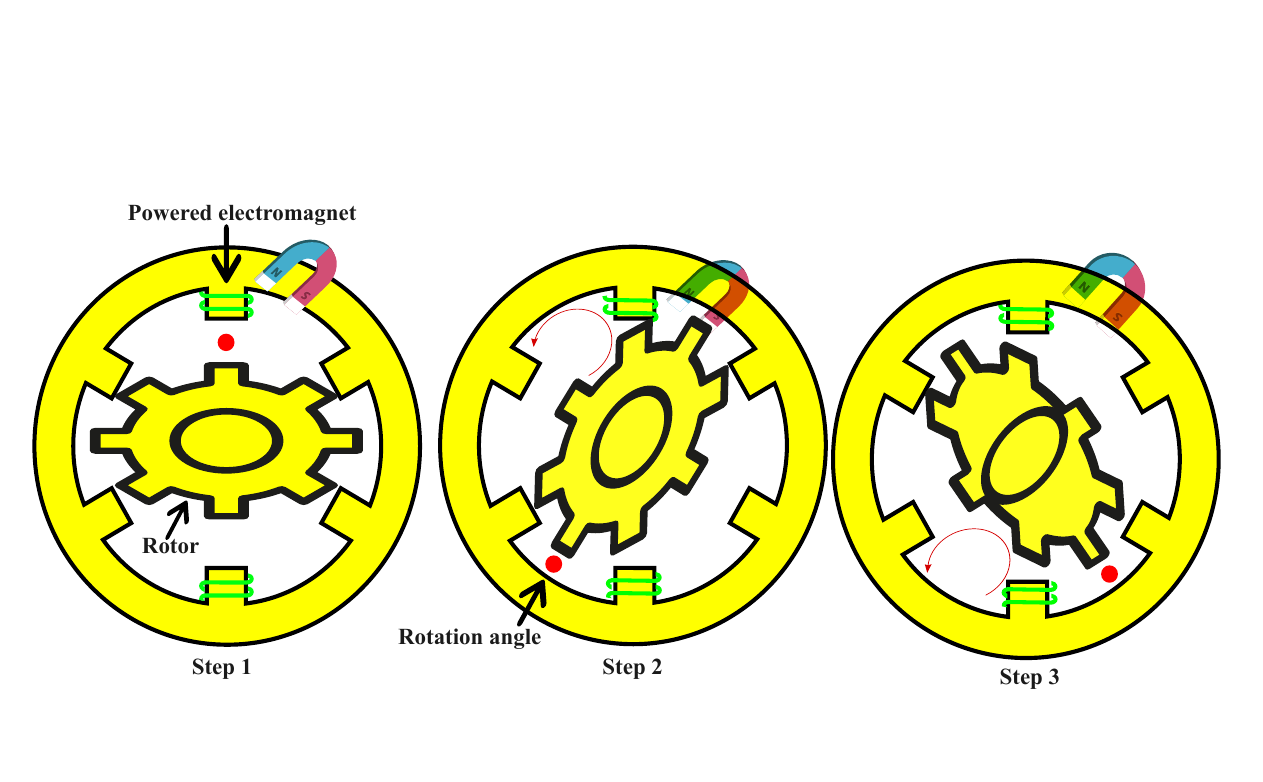} \vspace{-3mm}
    \caption{Exploded view of a common stepper motor used in 3D printers}
    \vspace{-3mm}
    \label{fig:your-image}
\end{figure}

The electromagnets are initiated as the stepper motors receive current by PWM technology. With the implementation of the rotor, the electromagnet power is transmitted to the turning movement of the bearings and spacers. Connecting this rotation movement to the 3D printer by belts and long threaded screws, the nozzle will gain the power to move in a three-dimensional platform. As depicted in Figure 7, the stepper consists of a stator, which generates a magnetic field by the usage of coils, which send currents to the rotor. This helps with accurate positioning and alignment of each movement. 

\begin{figure}[h]
    \centering
    \includegraphics[width=\linewidth]{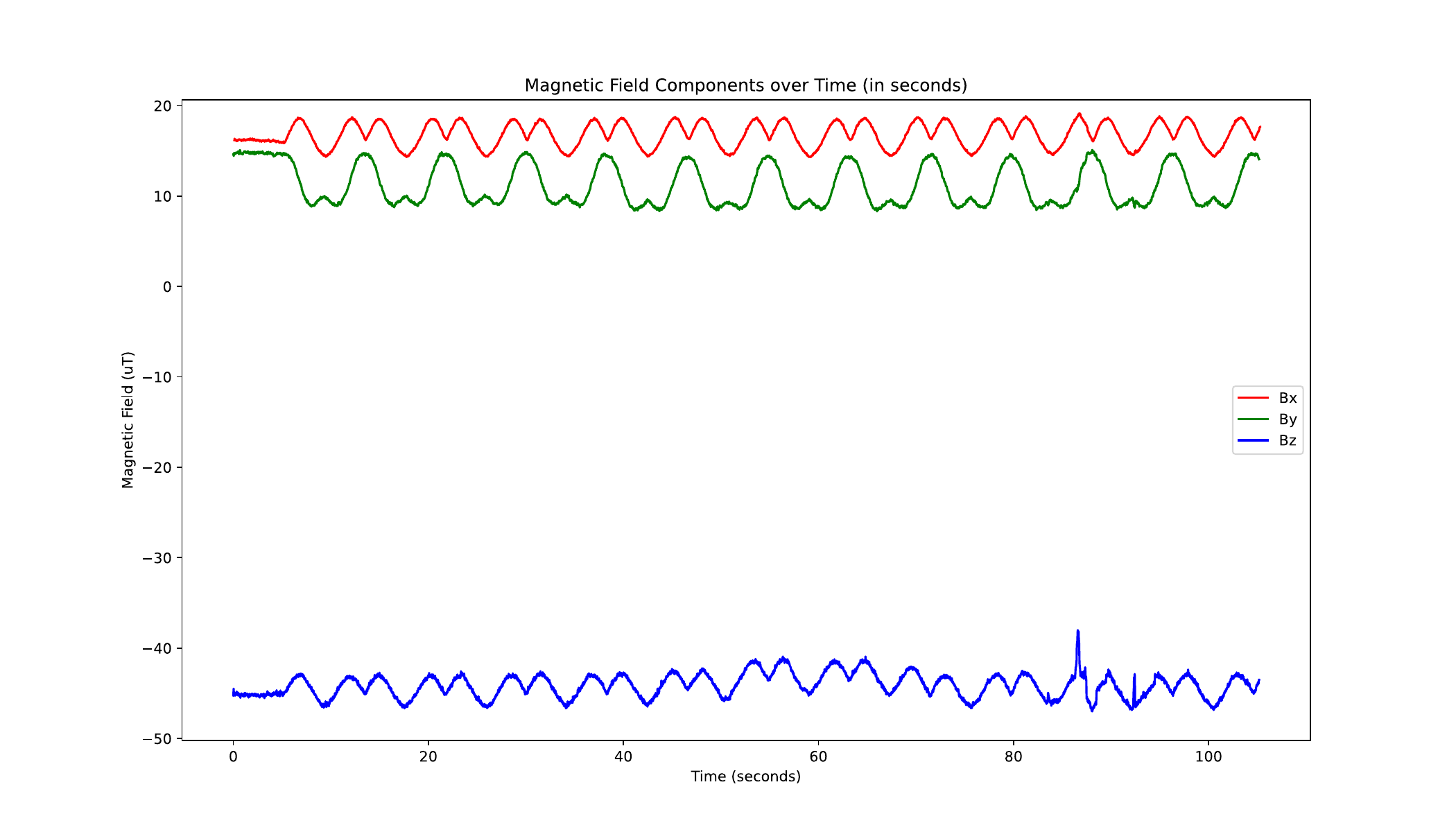} \vspace{-3mm}
    \caption{Magnetic data while the nozzle moves in x-axis} \vspace{-3mm}
    \label{fig:your-image}
\end{figure}

Knowing how steppers function, we used \cite{Samsung} to record magnetic data while the 3D printer only printed on the x-axis. We designed and sent a specific G-code through the slicing tool \cite{Cura}. The data was collected at 100Hz in the unit of micro-Tesla (µT). As is apparent in Figure 8, the pattern of the nozzle moving in the x-axis for X and Y magnetic fields was interestingly similar. This shows how the magnetometer received magnetic fields while the nozzle was in a left-right movement. As the nozzle gets closer to the left, where the smartphone was positioned, the magnetic field grows and picks at close to 20 µT, but as it moves to the right, further to the smartphone's built-in sensors, the magnetic field will become weaker. The blue wave also shows the z-axis movement, which is negative as obvious since we did not have any z-axis movement.

To make the data effective and classified, we use feature extraction as we did for acoustic data. This is considered as a pre-processing step for further analysis of the dataset. As to help the model to learn the central tendency better, we used Formula (8).  It assist a single value that represents the center of the dataset.

\begin{equation}
\text{Mean} = \frac{1}{N} \sum_{i=1}^{N} x_i
\end{equation}

Standard deviation (Std Dev) is applied because it detains how the spread of the data is. We harnessed Formula (9), which classifies different movements of a 3D printer.

\begin{equation}
\text{Std Dev} = \sqrt{\frac{1}{N} \sum_{i=1}^{N} (x_i - \text{Mean})^2}
\end{equation}

Skewness highlights whether data points are more concentrated on one side of the mean. A positive skewness indicates a right-tailed distribution (more values are concentrated on the left), and a negative skewness indicates a left-tailed distribution (more values are concentrated on the right).

\begin{equation}
\text{Skewness} = \frac{\frac{1}{N} \sum_{i=1}^{N} (x_i - \text{Mean})^3}{\left( \frac{1}{N} \sum_{i=1}^{N} (x_i - \text{Mean})^2 \right)^{3/2}}
\end{equation}

The kurtosis Formula (11) function calculates the tailedness or rocketedness of a distribution analogized to the normal distribution. 

\begin{equation}
\text{Kurtosis} = \frac{\frac{1}{N} \sum_{i=1}^{N} (x_i - \text{Mean})^4}{\left( \frac{1}{N} \sum_{i=1}^{N} (x_i - \text{Mean})^2 \right)^{2}} - 3
\end{equation}

\section{Acoustic and Magnetic Models}

Upon inspecting the 3D printer movements while printing an object, we found three key movements, as illustrated in Figure 9. By collecting distinct acoustic and magnetic data during each key stage and doing feature extraction, the 3D printer's initial IP will be at risk of being accessible. First, the nozzle moves either vertically or horizontally. If it moves vertically, then it is on the X-axis, moving left and right. However, if it moves horizontally, the nozzle moves up and down. The next step, header movement, determines if the nozzle is printing or aligning to position at the right spot and then starts printing. The speed of the motor can detect the major difference in printing or aligning as it aligns with maximum speed but slows down while printing to prevent any string act or bad quality printed shape.
\begin{figure}
    \centering
    \includegraphics[width=0.58\linewidth]{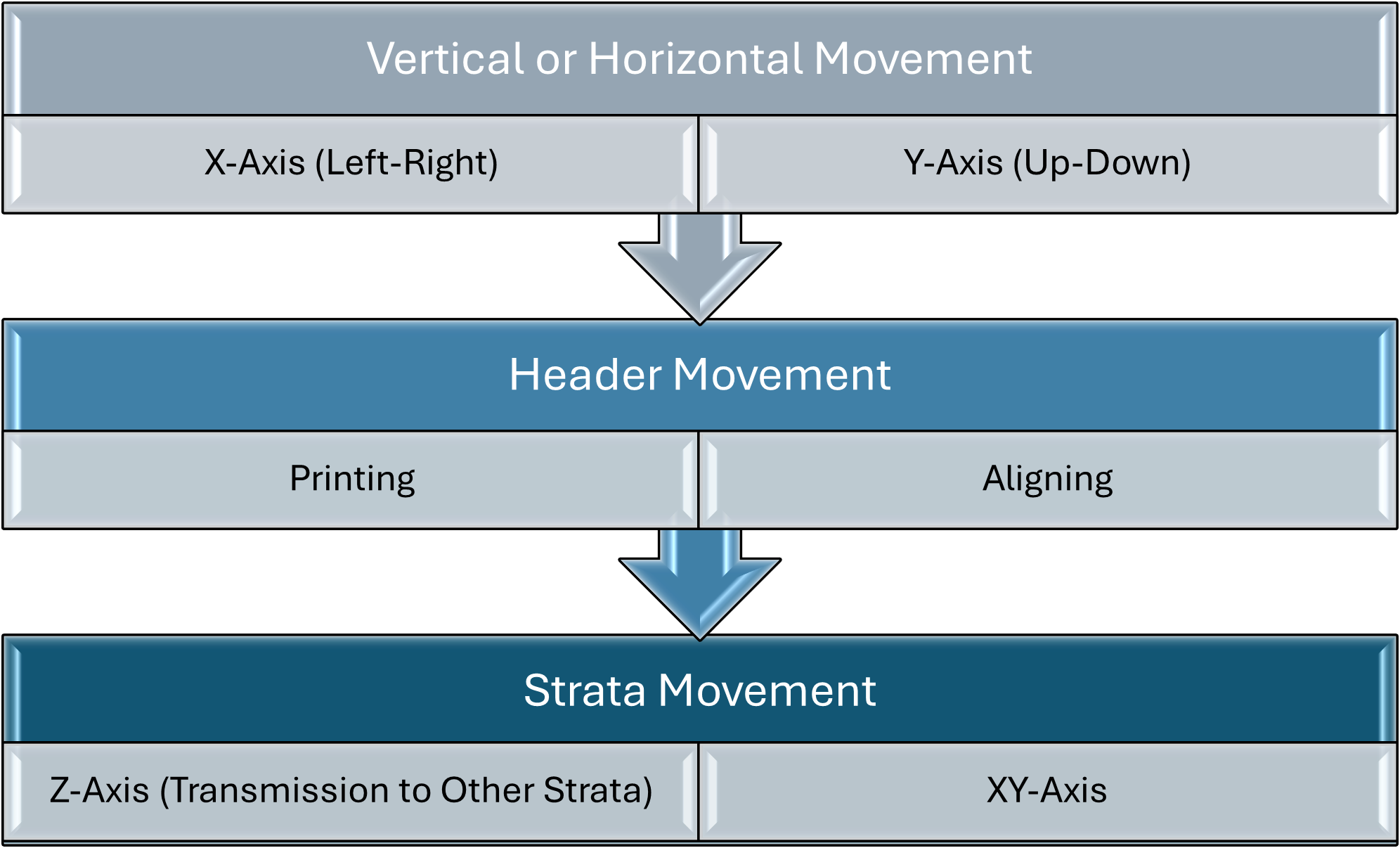}
    \caption{Taxonomy of the chronological order in 3D printers}
    \vspace{-3mm}
    \label{fig:enter-label}
\end{figure}

\subsection{Analysis and Results}
This section outlines our approach to identifying precise movement patterns in 3D printers. We employed Gradient Boosted Decision Trees trained on both acoustic and magnetic data. Using a Samsung smartphone equipped with sensors and strategically placed microphones, we captured magnetic data (100 µT) and acoustic data (dB) for thorough analysis. Data pre-processing included Gaussian filtering and segmenting signals into 100 ms frames, which were then organized into distinct training and testing datasets.

\subsection{Test Setup}

Having detected each specific movement mechanism in 3D printers, we formulated different models using Gradient Boosted Decision Trees to train the acoustic and magnetic data. The data collection setup involved positioning a smartphone \cite{Samsung}, equipped with built-in sensors, including an accelerometer, gyroscope, and magnetometer, alongside primary and secondary microphones capable of features like audio zoom and directional recording.
\begin{figure}
    \centering
    \includegraphics[width=0.60\linewidth]{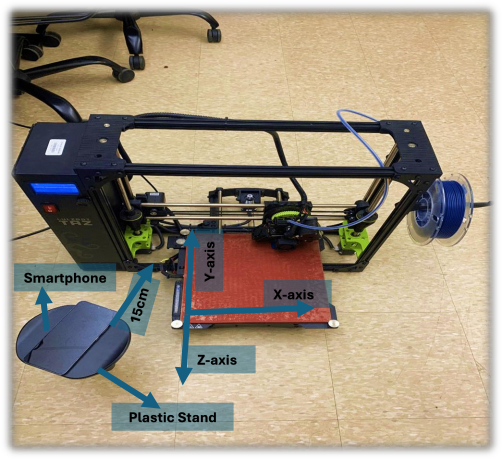}
    \caption{Experiment setup}  \vspace{-3mm}
    \label{fig:enter-label}
\end{figure}
Placed 15 cm away from the 3D printer at a 45-degree angle, the smartphone captured data optimally as shown in Figure 10.

The magnetic data, measured in 100 \(\mu\mathrm{T}\), and the acoustic data, recorded in dB, were both collected for detailed analysis of the printer's movements and performance. Initially, the side-channel data underwent smoothing using a Gaussian filter. Subsequently, the signal was segmented into frames, each lasting 100 ms. These segments were then categorized into training and testing sets tailored to the requirements of various models, ensuring robust analysis and prediction capabilities.

\subsection{Results}
Applying feature extractions to the collected data mentioned in Sections 4.1 and 4.2 and smoothing the data with the Gaussian filter, we were able to test our models and reach high accuracy in predicting which direction the nozzle is moving, as shown in Figure 11.

\begin{figure}[h]
    \centering
    \includegraphics[width=0.96\linewidth]{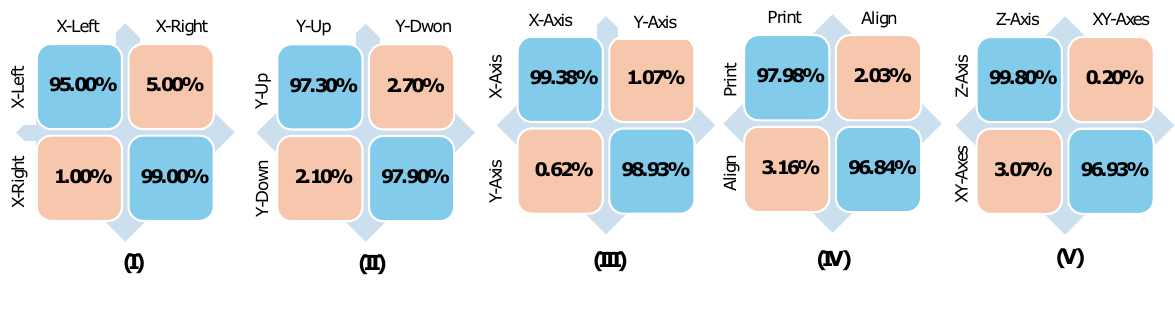}
    \vspace{-3mm}
    \caption{Model's accuracy per each movement}  \vspace{-4mm}
    \label{fig:your-image}
\end{figure}

Figures 11 (I) and 11 (II) show the vertical and horizontal models, which determine whether the printer operates in the X-left, X-right, or Y-up, Y-down plane. The training set comprises 1000 magnetic frames for each step, and the testing set includes a total of 3000 magnetic frames. This model can distinguish between the nozzle moving on either the X or Y axis and then detect if it is moving on the left and right axis or up and down. As apparent, the model successfully predicted 99.00\% of the movement while the nozzle was moving and printing in the X-left direction. 

Figure 11 (III) depicts the model's accuracy on diagnosing if the nozzle is moving in X or Y planes. This model is critical as it provides the foundation for knowing the movements in Figures 11 (I) and 11 (II). The training set comprises 1000 magnetic frames for each category, and the testing set includes a total of 3000 magnetic frames. This model effectively reached an average accuracy of 99.49\%. The high accuracy of this model stems from the fact that the magnetic data emitted while moving the nozzle in the X or Y plane was quite strong, so more precise data could be recorded by the smartphone magnetometer.

To firmly determine if the nozzle is actually printing or just positioning on the platform, we developed a model in Figure 11 (IV). Since the term speed plays a critical role in detecting the movements while the header is adjusting or printing, we used only acoustic data here and removed the magnetic data for the purpose of having tailored data with low deviation rate. The average accuracy for this model hit 97.25\%. Achieving high accuracy in nozzle detection, whether it is positioning or printing, is a complex scenario due to a combination of mechanical, physical, and software-related factors. Even small errors in any of these areas can add up and lead to noticeable inaccuracies in the final print.
\begin{figure}[h]
    \centering
     \vspace{-3mm}
    \includegraphics[width=\linewidth]{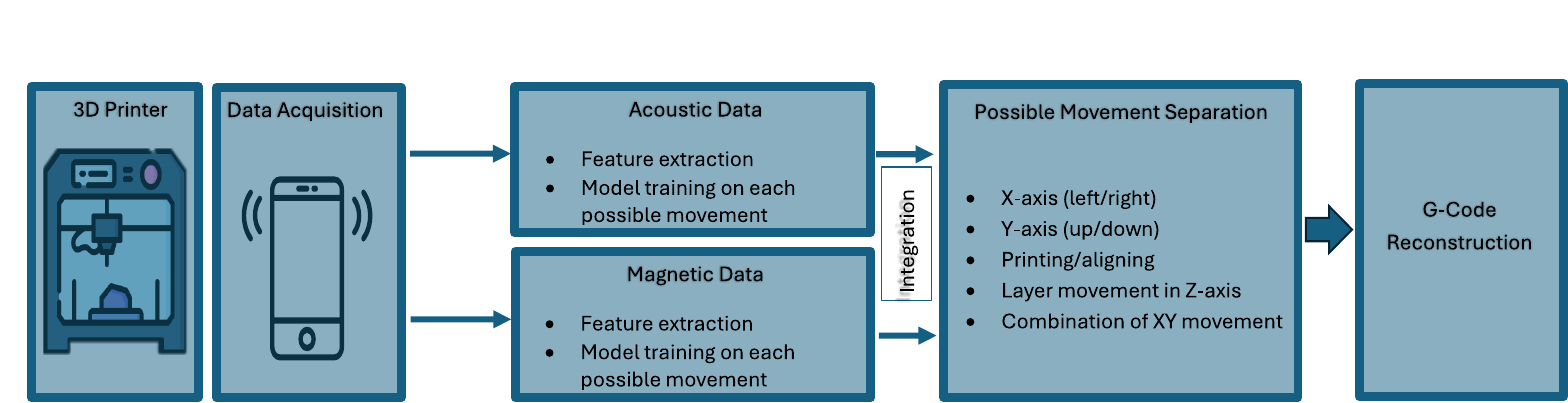}
    \caption{Overview of the stages to reconstruct G-code}  \vspace{-3mm}
    \label{fig:your-image}
\end{figure}

We developed a specialized model to distinguish between the Z-axis and XY-axis movements of the nozzle. Recognizing the distinct acoustic signatures associated with vertical versus horizontal movements, we focused inclusively on acoustic data and magnetic data. This streamlined approach resulted in a tailored dataset with minimal deviation, contributing to an average accuracy of 97.44\% as showing in Figure 11 (V).

\section{Real Environment Implementation}

To evaluate our models in a real-world examination, we decided to print a shape that involves every movement of the 3D printer for 3 layers: a square with dimensions of 1cm * 1cm. We followed steps in Figure 12. We positioned the smartphone near (within 15cm) the 3D printer and started to collect acoustic and magnetic data simultaneously, as shown in Figure 10, while the printer was executing our specified G-code for the shape of the square. Once the printing process was finished, we did the feature extraction steps on the data and used a Gaussian filter to smooth the dataset. We inserted this dataset as an input to the model we trained. With the implementation of data simulation from G-code instructions on Python, we were able to reconstruct the initial G-code of the printer with high accuracy in the final shape. As depicted in Figure 13, the overall square commands are successfully reconstructed with only some modifications in the axis and speed of the nozzle. Also, as it is apparent in Figure 14, there are some edge lay-offs and or increases in length for the reconstructed G-code. 
\begin{figure}[h]
    \centering
    \includegraphics[width=0.75\linewidth]{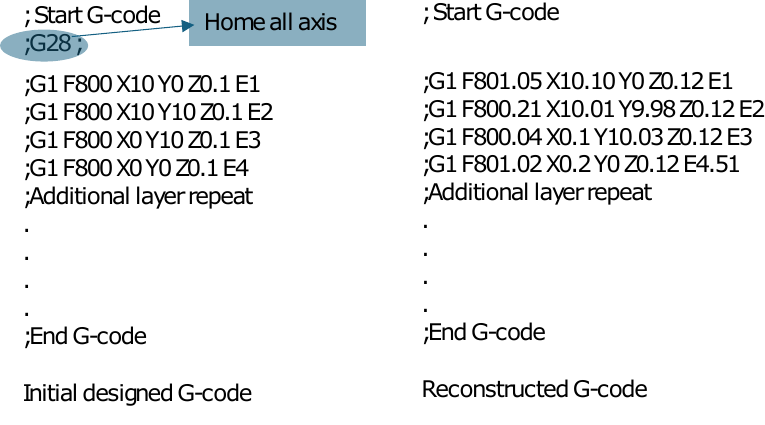} \vspace{-3mm}
    \caption{Comparison of the initial G-code with the reconstructed one by the use of side-channel attacks} \vspace{-3mm}
    \label{fig:gain_vs_time}
\end{figure}

\begin{figure}[h]
    \centering
\includegraphics[width=0.8\linewidth]{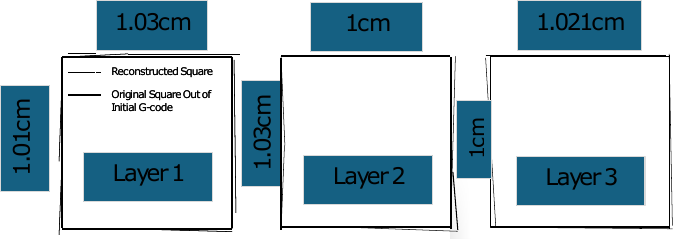}
    \caption{The visual comparison of the reconstructed and original square shape out of magnetic and acoustic data} \vspace{-4mm}
    \label{fig:your-image}
\end{figure}

MTE rate on this square shape reconstruction was 4.47\% which typically relates to a statistical measure used to assess the accuracy or bias of a forecasting or prediction model. In this case, the lower the MTE is, the higher the precision is on the reconstructed shape. It would generally involve calculating the average difference between predicted values and actual values over a dataset \cite{10.1145/2976749.2978300}, reflecting whether the model tends to underpredict or overpredict systematically. This metric helps in understanding the overall directional bias of the model's predictions.

\subsection{Discussion}

In this section, we discuss various limitations, including the distance between the smartphone and the 3D printer, differences in speed, equal loads on stepper motors, and background noise. Additionally, we compare our overall results with the related works that deployed relatively the same methods as ours to reconstruct the G-code. 

Distance has posed a significant constraint in the realm of side-channels \cite{heyszl2013strengths}. As demonstrated in Table 1, distance profoundly influences both the MTE and the G-code reconstruction, which is at the final stage of IP access. In the same way, there is a direct relationship between the complexity of the object that is designed to be printed with the overall MTE rate; The more complicated the object is, the lower the MTE rate would be as intricate object demands high speed of nozzle movement and nuance extrusion on the nozzle, which leads to the lower accuracy of collecting data. This is primarily due to the diminishing signal strength received by smartphones at greater distances. Consequently, the training of models is hindered by increased data variability, posing challenges for accurate movement prediction. Considering this limitation, our experiment and methods used reached high accuracy in reconstructing the G-code with low rates of MTE in collecting data with the smartphone within three different distances of 15cm, 20cm, and 30cm.

\begin{table}[ht]
    \centering
    \begin{tabular}{{p{5cm}c}}
        \toprule
        Distance (cm) & MTE Rate (percentage) \\
        \midrule
        20 cm & 5.10 (\%) \\
        30 cm & 6.09 (\%)\\
        \bottomrule
    \end{tabular}
    \caption{MTE rates at different distances} \vspace{-3mm}
    \label{tab:mte-rates}
\end{table}

However, one of the key points of the attack scenario tested in this study is the non-intrusive nature of the attack. This adds to the stealthiness of the experiment as, in opposition to traditional attack overviews, there were apparent modifications and intrusive actions on the printing process to collect the data. Although the distance between the smartphone and the 3D printer directly affects the accuracy and overall performance of the attack, still the same object can be reproduced by reconstructing the G-code with minor differences in the quality and specifications of the final object. It is noted that if the magnetic field is beyond the coverage range, smartphones will be unable to detect magnetic data. This limitation can lead to significant inaccuracies in object reconstruction.

Future work could explore using multiple smartphones and sensor fusion to further extend the attack distance. Additionally, it might be possible to investigate conducting such attacks without a direct line of sight to the victim device. 
Since acoustic and magnetic side channels do not always require a clear line-of-sight transmission between the sensor and the target object or victim \cite{liu2020maghacker,tu2023auditory}, adversaries could better conceal their devices in certain non-line-of-sight scenarios in real-world attack settings.

In addition, speed and consistent loads on stepper motors present additional limitations. The rapid alignment speeds typical in 3D printers result in unreliable magnetic and acoustic data collection during these fast operations. This rapid nozzle movement, driven by the high-speed actions of the steppers, complicates the accurate capture of data within short time frames. Similarly, when stepper motors operate under similar loads, distinguishing between them becomes challenging due to their production of magnetic and acoustic data that lacks significant distinctions. Training models under these conditions become particularly difficult due to the absence of clear differentiation between motor behaviors.

\begin{table}[h]
\centering
\begin{tabular}{lcccc}
\toprule
\textbf{Study} & \textbf{Metric} & \textbf{Value} & \textbf{Metric} &\textbf{Value} \\
\midrule
Our Study & Avg. Accuracy & 98.80\% & MTE & 4.47\%\\
\cite{10.1145/2976749.2978300} & Avg. Accuracy & 94.97\% & MTE & 5.87\%\\
\cite{al2016acoustic} & Avg. Accuracy & 78.35\% & MTE & Not Mentioned\\
\cite{chhetri2017confidentiality} & Avg. Accuracy & 86\% & MTE & Not Mentioned\\
\cite{Faruque2016AcousticSA} & Avg. Accuracy & 98.55\% & MTE & Not Mentioned\\
\bottomrule
\end{tabular}
\caption{Comparison of accuracy metrics between different studies} \vspace{-3mm}
\label{table:comparison}
\end{table}

Background noise presents another challenge in real-world IP attacks. While magnetic data remains largely unaffected, background noise significantly impacts the accuracy of acoustic data used to record stepper speed and movement. Despite efforts to mitigate this interference through feature extraction and Gaussian filters to reduce and smooth unwanted noise, the overall accuracy of data analysis and model predictions is inevitably influenced.

As shown in Table 2, the average accuracy and MTE rate of each related study have been summarized. The nearest study which had a high accuracy to ours was in \cite{Faruque2016AcousticSA}. Some of the studies did not mention the MTE rate once the G-code was reconstructed; however, \cite{Faruque2016AcousticSA} highlighted a MAPE rate of 3.13\% instead of MTE.

\section{Conclusion and Future Work}

AMs, specifically 3D printers, have become pervasive throughout the globe and are being used in different sectors. The importance of 3D printers in this world has caused researchers to protect the 3D printer's IPs and the G-code. However, since stepper motors produce magnetic field and acoustic sound, by recording and analyzing the data close to the printer, attackers can reconstruct the G-code and modify it as their intent. In this study, we explained how stepper motors generate data and outlined different movements in 3D printers. We trained a model using Gradient Boosted Decision Trees to diagnose the nozzle movement on the platform. Reaching high accuracy in predicting the movements, we implemented data simulation from G-code instructions on Python to gain access to the initial G-code. Having tested this method on a real-world object, we successfully reconstructed the G-code with an MTE of 4.47\%. In our future works, we plan to collect data from a greater distance using two smartphones positioned at different angles. This will help us investigate whether attackers can record data and regenerate the G-code from further away. Also, we will test our models on more complicated objects with more layers and nuance movements of the nozzle to check if the accuracy remains high. We hope the outcomes in this paper help to secure the valuable IPs of 3D printers. 

\section*{Acknowledgment}

This work was supported in part by the U.S. National Science Foundation under grants OIA-1946231, CNS-2117785, and CNS-2231682.

\bibliographystyle{splncs04}
\bibliography{AmirCite}

\end{document}